\documentstyle[own_sngl]{article}
\def\simless{\mathbin{\lower 3pt\hbox
   {$\rlap{\raise 5pt\hbox{$\char'074$}}\mathchar"7218$}}}   % < or of order
\def\simgreat{\mathbin{\lower 3pt\hbox
   {$\rlap{\raise 5pt\hbox{$\char'076$}}\mathchar"7218$}}}   % > or of order
\slugcomment{{\em  MNRAS, in press 9.06.98}}
\lefthead{}
\righthead{}

\begin{document}
%***************************************************************
%
\title {The dynamical evolution of stellar super-clusters}

\author {Pavel Kroupa\\
\medskip
\small{Institut f{\"u}r Theoretische Astrophysik, Universit{\"a}t
Heidelberg, \\
Tiergartenstr. 15, D-69121 Heidelberg \\
e-mail: pavel@ita.uni-heidelberg.de\\}} 

%%%%%%%%%%%%%%%%%%%%%%%%%%%%%%%%%%%%%%%%%%%%%%%%%%%%%%%%%%%%%%%%%%%%%%
\begin{abstract}
Recent images taken with the Hubble Space Telescope (HST) of the
interacting disk galaxies NGC~4038/4039 (the {\it Antennae}) reveal
clusters of many dozens and possibly hundreds of young compact massive
star clusters within projected regions spanning about~100
to~500~pc. It is shown here that a large fraction of the individual
star clusters merge within a few tens to a hundred~Myr.  Bound stellar
systems with radii of a few hundred~pc, masses~$\simless10^9\,M_\odot$
and relaxation times of $10^{11}-10^{12}$~yr may form from these.
These spheroidal dwarf galaxies contain old stars from the pre-merger
galaxy and much younger stars formed in the massive star clusters, and
possibly from later gas-accretion events.  The possibility that star
formation in the outer regions of gas-rich tidal tails may also lead
to super clusters is raised.  The mass-to-light ratio of these objects
is small, because they contain an insignificant amount of dark matter.
After many hundred~Myr such systems may resemble dwarf spheroidal
satellite galaxies with large apparent mass-to-light ratia, if tidal
shaping is important.
\end{abstract}
%%%%%%%%%%%%%%%%%%%%%%%%%%%%%%%%%%%%%%%%%%%%%%%%%%%%%%%%%%%%%%%%%%%%%%

%%%%%%%%%%%%%%%%%%%%%%%%%%%%%%%%%%%%%%%%%%%%%%%%%%%%%%%%%%%%%%%%%%%%%%
%%%%%%%%%%%%%%%%%%%%%%%%%%%%%%%%%%%%%%%%%%%%%%%%%%%%%%%%%%%%%%%%%%%%%%

%%%%%%%%%%%%%%%%%%%%%%%%%%%%%%%%%%%%%%%%%%%%%%%%%%%%%%%%%%%%%%%%%%%%%%
\section{Introduction} 
\label{sec:intro}
In the inner regions of the tidal arms of the interacting galaxy pair
NGC~4038/4039, Whitmore \& Schweizer (1995) observe with the HST many
hundreds of young compact star clusters. Using the re-furbished HST,
Whitmore et al. (1998) count thousands of massive star clusters with
dimension of a few~pc\footnote{Images are available at
http://\-oposite.stsci.edu/\-pubinfo/pr/97/34/}. The brightest and
bluest of the star clusters are probably younger than 10~Myr. Many
hundreds of similar compact star clusters with ages of a few
hundred~Myr are observed in the merger remnant galaxy NGC~7252 by
Miller et al. (1997).  The measured luminosity function of these
clusters rises steeply with decreasing luminosity, and shows no
turnover at the completeness limit. The luminosity function appears to
evolve to the form of the old globular clusters as the average nuclear
age of the star cluster system increases\footnote{A compilation by
B.C. Whitmore of cluster luminosity functions in mergers of different
ages is to be found at
http://oposite.stsci.edu/pubinfo/pr/97/34/images/HistogramsS.GIF}. This
is to be expected as a natural consequence of internal dynamical
evolution and tidal destruction of the least massive star clusters.

This impressive observational evidence strongly suggests that globular
clusters form in interacting galaxies.  The convergent gas flows in
colliding gas-rich galaxies lead to localised gas pressures that are
very high. Such regions can become gravitationally bound if the gas
mass is sufficiently large, favouring the formation and survival of
massive clusters (Elmegreen et al. 1993; Elmegreen \& Efremov 1997).
This is corroborated by the presence of young globular clusters in the
Large Magellanic Cloud (LMC), which is interacting mildly with the
Milky Way (see e.g. Westerlund 1997).  Of special interest in the
present context is that some of the LMC clusters appear to form
interacting binary systems (e.g. Vallenari, Bettoni \& Chiosi 1998).
It is thus likely that the compact clusters observed in the Antennae
galaxies evolve to globular clusters once the massive stars have
faded.

A particular stunning discovery made with the new HST images
is that the young compact star clusters are themselves clustered. The
super-clusters contain at least a few dozen bright star clusters
within a projected region no larger than a few hundred~pc, and are
centrally concentrated with a core radius of approximately~100~pc or
less. Some super-clusters appear to contain hundreds of massive star
clusters, with unresolved central regions.  It is interesting to note
that some super-clusters also appear to be clustered in groups of a
few.

The aim of this paper is to investigate the likely fate of such
super-clusters, and their possible relationship to spheroidal dwarf
galaxies. Also, the dynamical evolution of globular clusters in
spheroidal dwarf galaxies subject to tidal damage is briefly
discussed.  The dynamical state of the super-clusters is considered in
Section~\ref{sec:state}, and Section~\ref{sec:pop} contains an
estimate of their stellar populations.  N-body simulations from the
literature, which are described in Section~\ref{sec:sim}, are applied
to this problem in Section~\ref{sec:dynev}.  The results are discussed
in Section~\ref{sec:disc}, and the conclusions are given in
Section~\ref{sec:concl}.

\section{Dynamical Stability} 
\label{sec:state}
The new HST images suggest that most of the gas has already been
cleared from the visible super-clusters. Only few of the compact star
clusters that are within the super-clusters appear reddened. Whitmore
\& Schweizer (1995) estimate that only about 0.1~Myr are required for
a region of radius 10~pc to be cleared from gas and dust.  If gas
expulsion had unbound the compact star clusters then they should
disperse within roughly one crossing time, which is less than 1~Myr
for a star cluster containing $\simgreat10^5$ stars within a few~pc,
implying a velocity dispersion of about 10~pc/Myr.  Since most of the
star clusters are older than 5~Myr (Whitmore \& Schweizer 1995) and
are still $<10$~pc in extend, it follows that they are most
probably bound objects.

The time-scale for evaporation of the compact clusters due to internal
dynamical evolution is at least 1~Gyr, while it is longer than a
Hubble time for typical globular clusters.  Stars more massive than
$8\,M_\odot$ are lost from the cluster through supernova explosions
within about 50~Myr.  Their loss will not unbind the cluster because
they contribute only about 6-10~per cent of the cluster mass for a
solar neighbourhood stellar mass function (see below). Stars more
massive than $2\,M_\odot$ evolve off the main sequence with
significant mass loss within about 500~Myr. They contribute about
25~per cent to the cluster mass, so that their loss will also not
unbind the cluster.

It is unlikely that the compact star clusters comprising the
super-clusters fly apart owing to a large internal velocity
dispersion.  If the groups of clusters were unbound and expanding with
velocities typical for the stellar kinematics of the parent galaxy
(i.e. roughly 30~pc/Myr) then, given their ages of roughly 10~Myr,
they ought to be either much more dispersed than they are (600~pc in
diameter), or they would have had to be formed in a point-like region
containing the entire mass of the super-cluster in order to have the
dimensions observed now.  Both appear unlikely, especially since there
are at least about ten super-clusters with similar extensions visible
at different locations in the images. 

The brightest individual compact star clusters imaged by Whitmore et
al. (1998) and Whitmore \& Schweizer (1995) have an integrated
absolute V-band magnitude of $M_{V,{\rm cl}}\approx-15$. A lower limit
cannot at present be determined, but the luminosity function of the
young star clusters is a power-law, $\phi\propto L^{-1.78\pm0.05}$
(Whitmore \& Schweizer 1995), where $L$ is the integrated luminosity
of an individual cluster, and $\phi dL$ is the number of clusters in
the luminosity range $L$ to $L+dL$. The luminosity function shows no
turnover down to the completeness limit of $M_{V,{\rm cl}}\approx 10$.

An O-type star with a mass of $m=40\,M_\odot$ has $M_V\approx -5.6$,
so that the above range on $M_{V,{\rm cl}}$ roughly corresponds to
clusters of 5800~O stars down to 58~O stars. If the initial mass
function is similar to what is seen in the solar neighbourhood (Scalo
1986 for $m\simgreat1\,M_\odot$; Kroupa, Tout \& Gilmore 1993 for
$m\simless1\,M_\odot$), then each O~star is associated with roughly
2300 stars less massive than $1\,M_\odot$. The star clusters then
consist of roughly $1.3\times10^7$ down to $1.3\times10^5$ stars,
which corresponds to cluster masses in the range $10^7\, M_\odot$ down
to $10^5\,M_\odot$ for a mean stellar mass of ${\overline
m}=0.5\,M_\odot$.

A super-cluster containing tens or hundreds of such clusters can thus
have a total mass of $M_{\rm scl}\approx10^6-10^9\,M_\odot$.  This
mass is contained within a projected region with radius of roughly
200~pc.  The tidal radius is $r_{\rm t}\approx[M_{\rm scl}/(3\,M_{\rm
gal})]^{1/3}R$ (Binney \& Tremaine 1987), where $M_{\rm gal}$ is the
mass of the galaxy within the distance $R$ of the super-cluster from
the centre of the galaxy. The super-clusters lie at a distance of
typically $R\approx5$~kpc from the centre of NGC~4038 in the
Antennae. The tidal radius is thus $r_{\rm t}\approx90-900$~pc if
$M_{\rm gal}\approx6\times10^{10}\,M_\odot$, which is the mass within
5~kpc of a disk galaxy with a circular velocity of~220~km/s. This
estimate of $r_{\rm t}$ can only serve as a guide because the galactic
mass distribution within $R$ is highly uncertain given the seriously
perturbed state of both interacting galaxies.

The most massive super-clusters should thus be stable against tidal
disruption. The less massive ones will stretch along their orbits
leading to families of massive star clusters on very similar galactic
orbits.

\section{Stellar Population} 
\label{sec:pop}
There is no evidence for significant amounts of dark matter
distributed like the disk in the Galaxy (Gilmore, Wyse \& Kuijken
1989; Kuijken 1991).  Assuming the disks of the unperturbed galaxies
that now form the interacting pair in the Antennae were similar to the
Galactic disk, which is supported by the analysis of the mass-to-light
ratio and global stability of galactic disks by Syer, Mao \& Mo
(1998), it follows that the dark-matter content of the super-clusters
is insignificant: Only an insignificant amount of dark halo matter
will become bound to the super-clusters because it has a velocity
dispersion of the order of 100~km/s. This is explicitly shown in the
simulations of interacting galaxies leading to the probable formation
of tidal-tail dwarf galaxies (Barnes \& Hernquist 1992).

The super-clusters cover a spherical volume of about
$3\times10^7$~pc$^3$ for a radius of 200~pc.  Assuming a stellar
number density of $0.1$~stars~pc$^{-3}$, which is characteristic for
the solar neighbourhood, it follows that the volume should contain of
the order of $10^6-10^7$ disk stars.  Not all stars in this volume
will remain bound to the super-cluster.  The escape velocity, $v_{\rm
esc}=(2GM_{\rm scl}/r)^{1/2}$ (gravitational constant
$G=4.49\times10^{-15}\,{\rm pc}^3/(M_\odot {\rm yr}^2)$) from a
super-cluster with a mass $M_{\rm scl}=10^6-10^9\,M_\odot$ within a
radius $r=200$~pc is, for stars within that radius, $v_{\rm
esc}\simgreat7-200$~km/s.  Assuming the velocity dispersion is
$\sigma^*=30$~km/s (similar to what is observed in the solar
neighbourhood), and approximating the distribution of velocities by a
Gaussian function (Reid, Hawley \& Gizis 1995), about~13 (99)~per cent
of the stars have velocities relative to the local standard of rest of
less than~5 (78)~km/s. Thus, it is likely that about $10^5-10^7$ stars
in the respective volume will remain bound to the super-cluster, if
the super-cluster moves with the local standard of rest, and if the
super-cluster forms on a time-scale that is shorter than the crossing
time $t_{\rm cr}=2\,r/\sigma^*\approx13$~Myr, so that the velocities
of the older disc stars have no time to adjust to the deeper potential
well.  In the super-cluster, the proportion of bound old to new stars
may then approach a few tens~per cent. If, on the other hand, the
super-clusters form on a time-scale that is longer than $t_{\rm cr}$,
then the disk stars near the forming super cluster will accelerate
into the potential well, subsequently leaving it.

The older disc stars formed throughout the galaxy over its entire age
until the time of the formation of the super-clusters. The stellar
population in a super-cluster is therefore a mixture of stars formed
in the process of cluster formation and of stars belonging to the
galactic disk prior to the galaxy interaction.  The age and
metallicity distribution of the stars in super-clusters should thus be
complex, and some radial age and metallicity gradients are expected
because the older, metal deficient stars are less concentrated towards
the centre of the super-clusters.

\section{N-body Simulations} 
\label{sec:sim}
The crossing time of a super-cluster with diameter $d$ is
approximately $t_{\rm cross}=d/\sigma$, where $\sigma\approx
(G\,M_{\rm scl}/d)^{1/2}$ is the velocity dispersion in a spherical
system in virial equilibrium.  For $M_{\rm scl}=10^6-10^9\,M_\odot$
and $d=400$~pc, it follows that $\sigma\approx 3-100$~pc/Myr and
$t_{\rm cross}\approx130-4$~Myr.

The time-scale for cluster-cluster interactions is therefore
comparable to or longer than the present age of the young clusters
($\simless10$~Myr), and significantly shorter than the half-mass
relaxation time, $t_{\rm rh}({\overline m})\approx1$~Gyr, typical for
the individual massive star clusters.

Core collapse occurs within $2\,t_{\rm rh}({\overline m})-3\,t_{\rm
rh}({\overline m})$ for a cluster with a realistic stellar mass
function (Spitzer 1987; Inagaki 1985; Inagaki \& Saslaw
1985). However, massive stars with mass $m_{\rm m}$ will segregate
towards the cluster core on the equipartition time-scale, $t_{\rm
eq}\approx\left(m/{\overline m}\right)\,t_{\rm rh}$, if initially
there is no mass-segregation and stellar velocities are independent of
mass (Spitzer 1987, p.74). Under these conditions, and if $m_{\rm
m}\approx100\,{\overline m}$, some internal dynamical evolution will
result within 10~Myr. However, observations of young embedded clusters
show that the most massive stars are usually located at the cluster
centre, so that the stellar system will evolve on a time-scale longer
than $t_{\rm eq}$. For the present purpose, the internal dynamical
evolution of the individual star clusters through two-body relaxation
can thus be neglected over the dynamical evolution time-scale of the
super-clusters.  Furthermore, the thousands of O~type stars within the
super-clusters have swept out most of the remaining gas within less
than about~10~Myr (Section~\ref{sec:state}).

Collision-less simulations without gas of a super-cluster can thus be
performed to study its dynamical evolution on a time-scale that is
smaller than the relaxation time within an individual globular star
cluster.  Such simulations exist in the literature. Garijo,
Athanassoula \& Garc\'ia-Gom\'ez (1997, hereinafter GAG) report
simulations of clusters of 50~galaxies in order to investigate the
formation of centrally dominant galaxies.

\subsection{Scaling the problem}
\label{ssec:scale}
The simulations of GAG can be applied to the problem here by scaling
to the appropriate mass and length scales.  This means that the tidal
field of the merging galaxies is ignored. This is a reasonable
simplification in this pilot study in which the structural and
kinematical properties of an evolved super-cluster are to be
quantified.  A tidal field will be added in more detailed
investigations (Fellhauer \& Kroupa 1998).

The freedom of scale in the gravitational problem is expounded nicely
by Madejsky \& Bien (1993). The transformation equation between two
systems of units is $(L/L^*)^3 = (T/T^*)^2(M/M^*)$ (their equation~1;
$L= $ length, $T = $ time and $M = $ mass). Velocities scale as $V/V^*
= (L/T)(T^*/L^*)$.  The length-scale in the galaxy simulations of GAG
is defined by the Plummer radius of each galaxy, $L^*=6$~kpc, and the
mass-scale is defined by the mass of an individual galaxy,
$M^*=5\times10^{11}\,M_\odot$.  The unit of time in their simulations
is $T^*=140$~Myr. For the problem of interest here, the Plummer radius
of each star cluster is $L=6$~pc, the star-cluster mass is
$M=10^6\,M_\odot$ and the resulting unit of time is $T=3.1$~Myr.
Thus, a length of 100~kpc and a time-span of 1~Gyr in the galaxy
problem of GAG corresponds to a length of 100~pc and a time-span of
22~Myr in the super-cluster problem studied here.  Finally, a velocity
of $V^*=100$~km/s in their simulations corresponds to $V=4.5$~km/s in
the super-cluster problem here.

\subsection{Initial conditions}
\label{ssec:ic}
In the simulations of GAG the galaxies are initially represented as
identical Plummer spheres, each consisting of 900~particles. They
perform eight simulations using a tree-code adapted for a Cray
computer. The softening length is~1.5~pc in the units used here, and
the simulations are run for 4000~steps, i.e. 95~Myr in the units used
here.

The initial conditions are varied to give seven different
configurations of interest here (the eighth is the hollow
configuration in which there are no galaxies in the central region of
the galaxy cluster).  These are summarised in table~1 of GAG.  In the
following, a globular cluster is meant to mean a galaxy in their
simulations.  The distance of each of the 50~globular clusters to the
centre of the super-cluster is chosen randomly between~0 and~$r_{\rm
scl}$.  Four simulations with an outer radius of the super-cluster of
$r_{\rm scl}=900$~pc are run for initially collapsing super-clusters
with initially spherical, oblate and prolate shapes. Initially
spherical super-clusters in virial equilibrium are also simulated for
$r_{\rm scl}=600$~pc and 300~pc. The latter are referred to as the
``compact models''.

\section{Dynamical Evolution} 
\label{sec:dynev}
\begin{figure}[bp]
%\plotfiddle{f1.ps}{15cm}{0}{70}{70}{-200}{-40}
\epsscale{0.8}
\plotone{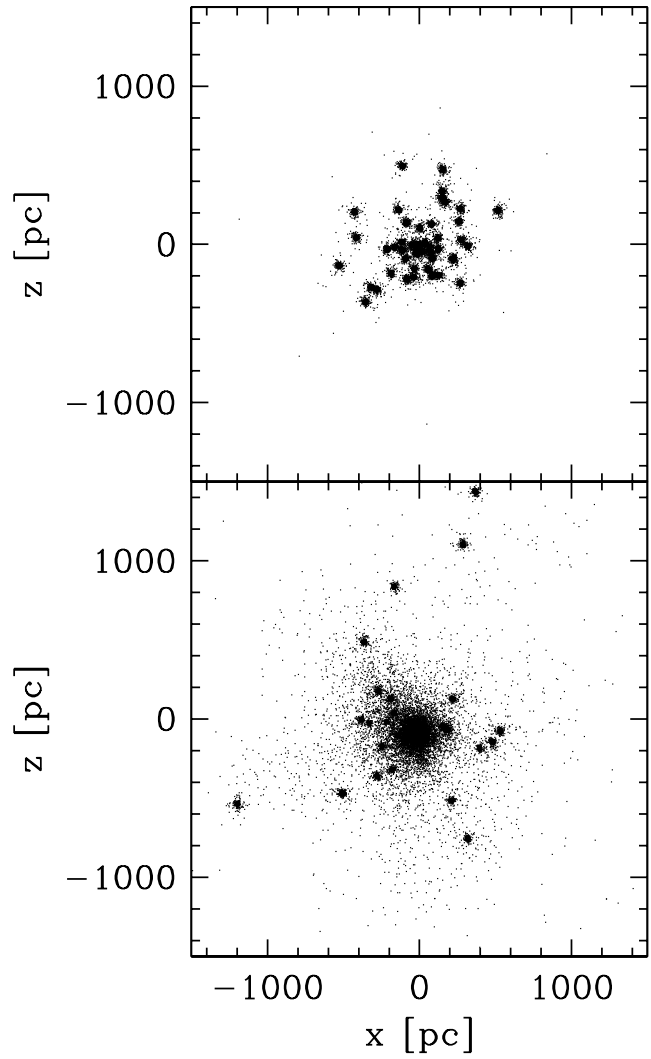}
\caption{
\label{fig:f1}
Two snap-shots from simulation~V of GAG. Initially (upper panel), the
super-cluster is in virial equilibrium.  Some of the clusters expand
outwards because they have positive radial velocities initially.  The
lower panel shows the resultant system after 95~Myr.}
\end{figure}

In all cases the globular clusters initially near the centre of the
super-clusters merge within a few~$10^7$~yr. Thereafter, individual
globular clusters continue to merge with the growing central stellar
system. Mass is also lost from the globular clusters through mutual
tidal forces, which leads to an extended stellar component occupying
the entire spherical volume within roughly $r_{\rm scl}$. Some
particles escape.  The configuration at the start of the simulation
with $r_{\rm scl}=600$~pc and after 95~Myr is shown in
Fig.~\ref{fig:f1}. Additional time-steps are displayed in fig.~1 of
GAG.
%%%%%%%%%%%%%%%%%%%%%%%%%%% Fig %%%%%%%%%%%%%%%%%%%%%%%%%
%%%%%%%%%%%%%%%%%%%%%%%%%%%%%%%%%%%%%%%%%%%%%%%%%%%%%%%%

The number of globular clusters decreases with time at different rates
depending on the initial configuration.  After 95~Myr less than 60~per
cent of the initial number of globular clusters remain in all seven
simulations. The cold collapse models and the compact models lead to a
survival fraction of~40~per cent by this time.  In the cold collapse
models, the mean radius of the surviving globular cluster system
decays but stabilises by about 62~Myr at the value of
roughly~300~pc. In the super-clusters that are initially in virial
equilibrium, the mean radius increases to~700-900~pc after~95~Myr
(fig.~2 in GAG), because the clusters on low-eccentricity and
long-period orbits survive preferentially.  

The mean velocity dispersion in the central object remains roughly
constant after about 40~Myr, and is determined by the initial
concentration, i.e. $r_{\rm scl}$. The velocity dispersion ranges
between~14 and~18~km/s.

Further details concerning the rate of growth of the mass of the
central stellar system, of its mean radius and the evolution of the
surviving globular clusters, of the axial ratio of the central stellar
system, of its density and velocity dispersion profiles can be found
in GAG.

The main interest here lies in the overall properties of the merged
super-cluster. Between about 40~and 60~per cent of the globular
clusters survive for 95~Myr. By this time, the centrally formed
objects have, approximately, three-dimensional Hernquist profiles with
half-mass radii in the range $r_{1/2}=45$ to~55~pc for the
super-clusters initially in virial equilibrium, and $r_{1/2}=73$
to~92~pc for the initially collapsing super-clusters
(Garc\'ia-Gom\'ez, private communication). Evolution of the half-mass
radius is slow, and it will not increase significantly after 100~Myr.
At 95~Myr, the centrally formed object has a projected surface density
profile that is flat within roughly the innermost 20~pc. Further out,
the surface density profile decreases with radius and approximately
follows a $r^{1/4}$ law with deviations that depend on the initial
configuration (figs.~16-18 in GAG). The projected velocity dispersion
profile is flat within the innermost 20~pc, and decreases roughly as a
power-law with increasing radius (fig.~20 in GAG). No significant
further evolution of these quantities is expected after 100~Myr.
Clearly, a time-dependent tidal field will alter both the three- and
two-dimensional density profiles.

\section{Discussion}
\label{sec:disc}
All initial conditions discussed here lead within~100~Myr to a central
object that has a half-mass radius of about 45--95~pc and consists of
about $10^8$ stars, and of surviving globular clusters on
low-eccentricity, long-period orbits.  The tidal radius would be, in
the case of the Antennae, about~300~pc (Section~\ref{sec:state}), so
that tidal shaping of the central object must be expected (Oh, Lin \&
Aarseth 1995).  The relaxation time (see Binney \& Tremaine 1987) of
this merged object is roughly $2-15\times10^{11}$~yr, taking for the
mean stellar mass ${\overline m} = 0.5\,M_\odot$ and for the number of
stars $10^7-10^8$. Practically, it is a collision-less system, and can
be called a dwarf galaxy.  It is spheroidal.  Deviations from
spherical symmetry, and the density profile, are determined by the
initial conditions as demonstrated by GAG, and by the tidal field
(Fellhauer \& Kroupa 1998).

Hereinafter such objects are referred to as {\it spheroidal dwarf
galaxies}, to distinguish them from the well-known dwarf-spheroidal
(dSph) and dwarf-elliptical (dE) galaxies. Spheroidal dwarf galaxies
are thus postulated to evolve from massive super clusters. They have
not been knowingly observationally identified yet.

However, the evolution of such a system at times longer than a few
hundred~Myr is interesting in the context of dE galaxies with nuclei,
and dSph satellites with globular clusters. A rough estimate of the
dynamical friction time, $t_{\rm dynfr}$ (see e.g. Binney \& Tremaine
1987), for a globular cluster with a mass of $10^5\,M_\odot$ in a
spheroidal dwarf galaxy with a mass of $5\times10^7\,M_\odot$,
suggests that a surviving cluster can sink to the centre within a
few~Gyr.  The accumulation of one or a few globular clusters at the
centre of such a stellar system should lead to a spheroidal dwarf
galaxy with a nucleus.

Future survival of the dwarf galaxy depends on its mass and its orbit
within the merging galaxy. Those not massive enough will fall apart in
the tidal field, contributing families of globular clusters on very
similar orbits.  The present configuration of the Antennae suggest
that approximately~200~Myr have passed since the first pericenter
passage (Barnes 1988). The two galaxies will probably suffer a second
pericenter passage within a few hundred~Myr, and will ultimately merge
when additional long tidal arms are flung out (Mihos, Dubinski \&
Hernquist 1998). The massive spheroidal dwarf galaxies formed now may
be ejected onto eccentric orbits with apo-galactica of tens of~kpc and
orbital periods of the order of a Gyr.  The final merged 'Antennae'
galaxy may evolve to an early-type disk galaxy after the ejected gas
re-settles in one plane (Hibbard \& Mihos 1995, Combes 1998), but
evolution to an elliptical galaxy is also possible.

The overall rate of sinking of globular clusters towards the centre of
the dwarf galaxy can be slowed if it looses stars in a tidal field.
Near each peri-galacticon, the dwarf is heated and expands as a result
of mass loss.  Any remaining globular clusters will then find
themselves at larger distances from the dwarf's centre, with $t_{\rm
dynfr}$ lengthened because of the dwarf's reduced mass and
density. The net result may thus be that globular clusters sink
towards the centre of a spheroidal dwarf galaxy increasing their
robustness against tidal removal from the dwarf. Mass loss near
peri-galacticon then essentially resets the dynamical friction clock
by leading to a general expansion of the dwarf. Thereafter the clock
ticks at a slower rate, and some clusters may never reach the centre
of the dwarf galaxy.

Viewed from this perspective, it is interesting to note that remnants
of globular clusters may be discernible as small-scale sub-structure
or as nuclei in dSph galaxies. The search for remnants of star
clusters in Galactic dSph satellites is difficult though because some
of them show significant sub-structure (Demers et al 1995; Irwin \&
Hatzidimitriou 1995). However, Ursa Minor contains a density
enhancement near its centre. This could be the remnant of a globular
cluster. Also, Fornax contains five globular clusters.

If a dwarf galaxy accretes a gas cloud that is on a similar orbit then
an additional period of star formation may take place within the
dwarf. If this were to occur then the gas is funnelled to the central
regions of the dwarf, possibly adding to or creating a nucleus (van
den Bergh 1986).  The resultant age and metallicity distribution of
the stars may thus show similarities to the observed distributions in
Galactic dSph satellites (see also Section~\ref{sec:pop}). These are
reviewed by Ferguson \& Binggeli (1994), Gallagher \& Wyse (1994),
Grebel (1997) and Da Costa (1997).

High-resolution imaging should show if the formation of super-clusters
is also a mode of star-formation in outer tidal arms.  That dwarf
galaxies may form in tidal tails has been surmised for some time
(e.g. Lynden-Bell 1976).  Evidence for this is reviewed by Duc \&
Mirabel (1998) and Kroupa (1998a). It is unclear though if such
gas-rich objects survive the star-formation episode as bound entities,
nor if they are sufficiently stable against tidal field destruction to
survive many orbits (see Kroupa 1998b).  If rich groups of globular
clusters form in these then, as shown here, bound spheroidal dwarf
galaxies may remain after gas expulsion, adding to the faint end of
the galaxy luminosity function.  Such tidal-dwarf galaxies are void of
dark matter, but may appear to be dark-matter dominated after
significant tidal shaping (Kroupa 1997; Klessen \& Kroupa 1998). These
issues are of relevance to the epoch of galaxy construction. If most
spiral galaxies formed at a red-shift of $1\simless z\simless2$ from merging
gas-rich sub-structures (e.g. Driver et al. 1998) in which some super
clusters are formed, then today's larger galaxies might naturally end
up with systems of spheroidal satellite galaxies.

\section{Conclusions} 
\label{sec:concl}
The clusters of many dozens and hundreds of massive star clusters
imaged in the inner parts of the tidal arms in the Antennae evolve on
a time-scale that is comparable to or somewhat longer then the present
age of the super-clusters, but significantly shorter than the
half-mass relaxation time of the individual massive star clusters. The
evolution may be approximated using a collision-less method, and leads
to the formation of a central stellar system that is more extended
than the individual star clusters by one to two orders of
magnitude. It has a mass of roughly $10^6-10^9\,M_\odot$, depending on
how populous the super-cluster was, and consists of stars formed
during the formation of the super-cluster and of old stars from the
parent galaxy, leading to radial metallicity and age gradients within
the object.  It's relaxation time is $10^{11}-10^{12}$~yr, so that it
may be called a spheroidal dwarf galaxy. It contains essentially no
dark matter apart from stellar remnants.

Tidal stability of the most massive young dwarf galaxies is likely,
but tidal modification at birth will be important.  If the dwarf
galaxy formed in the inner parts of a tidal arm during or after the
first galaxy--galaxy encounter, then it is feasible that it may be
ejected during the final merging encounter onto an eccentric orbit
with an apo-centric distance of tens of~kpc.  The dwarf galaxy will
subsequently evolve through periodic tidal modification and may appear
similar to some of the Galactic dSph satellites.

Since not all of the massive star clusters originally in the
super-cluster merge within the first few hundred Myr, and since they
are expected to evolve to globular clusters, it follows that some of
the more massive spheroidal dwarfs may contain a few globular
clusters, a few of which may sink to their centres contributing to or
producing a nucleus.  Tidally induced mass loss from the dwarf,
however, may compensate orbital shrinkage through dynamical
friction. In this case, some globular clusters may remain bound to the
dwarf without reaching its centre.  The globular clusters are younger
than the oldest stars in the dwarf.  Those stripped from the dwarf
will remain members of the tidal stream. Super clusters not massive
enough to remain bound in the tidal field will disassemble, producing
families of globular clusters on similar orbits.

The formation of stellar super-clusters may be an important mode of
star formation during galaxy assembly at $1\simless z\simless2$, and at
$z\approx0$ in tidal-tail dwarf galaxies which condense on highly
eccentric orbits at distances of many tens of~kpc from the parent
galaxies.  Such dwarfs are distinct from dwarf galaxies that formed
from the Hubble flow in that they contain essentially no dark matter.

%%%%%%%%%%%%%%%%%%%%%%%%%%%%%%%%%%%%%%%%%%%%%%%%%%%%%%%%%%%%%%%%%%%%%%
\acknowledgements 
\vskip 10mm
\noindent{\bf Acknowledgements}
\vskip 3mm
\noindent
I am grateful to Carles Garc\'ia-Gom\'ez for helpful and kind
responses to my requests, and I thank Ralph Andersen, Rainer
Spurzem and Holger Baumgardt for useful discussions. 

%%%%%%%%%%%%%%%%%%%%%%%%%%%%%%%%%%%%%%%%%%%%%%%%%%%%%%%%%%%%%%%%%%%%%%
% Reference List:
%

\clearpage
%%%%%%%%%%%%%%%%%%%%%%%%%%%%%%%%%%%%%%%%%%%%%%%%%%%%%%%%%%%%%%%%%%%%%

\begin{references}

\reference{} Barnes J.E., 1988, ApJ, 331, 699
\reference{} Barnes J.E., Hernquist L., 1992, Nature, 360, 715
\reference{} Binney J., Tremaine S., 1987, Galactic Dynamics,
      Princeton, New Jersey
\reference{} Combes F., 1998, in Formation and Evolution of
	Galaxies, eds O. Le Fevre, S. Charlot, Les Houches Series,
	Springer Verlag, in press (astro-ph/9804121).
\reference{} Da Costa G.S., 1997, in Stellar Astrophysics for the
      Local Group: A First Step to the Universe, Proceedings of the
      VIIIth Canary Islands Winter School, eds A. Aparicio and
      A. Herrero, (Cambridge: Cambridge University Press), in press.
\reference{} Demers S., Battinelli P., Irwin M.J., Kunkel W.E., 1995,
      MNRAS, 274, 491
\reference{} Driver S.P., Fern\'andez-Soto A., Couch W.J., Odewahn
	S.C., Windhorst R.A., Phillipps S., Lanzetta K., Yahil A., 1998, ApJ,
	496, L93
\reference{} Duc P.-A., Mirabel I.F., 1998, A\&A, 333, 813
\reference{} Elmegreen B.G., Kaufman M., Thomasson M., 1993, ApJ, 412,
      90
\reference{} Elmegreen B.G., Efremov Y.N., 1997, ApJ, 480, 235
\reference{} Fellhauer M., Kroupa P., 1998, in preparation 
\reference{} Ferguson H.C., Binggeli B., 1994, A\&AR, 6, 67
\reference{} Gallagher J.S., Wyse R.F.G., 1994, PASP, 106, 1225
\reference{} Garijo A., Athanassoula E., Garc\'ia-Gom\'ez C., 1997, A\&A,
      327, 930 (GAG)
\reference{} Gilmore G., Wyse R.F.G., Kuijken K., 1989, ARA\&A, 27, 555
\reference{} Grebel E.K., 1997, Reviews in Modern Astronomy, 10, 29
\reference{} Hibbard J.E., Mihos J.C., 1995, AJ, 110, 140
\reference{} Inagaki S., 1985, in Dynamics of Star Clusters, IAU
        Symp. 113, eds Goodman J., Hut P., Dordrecht, D.Reidel
	Publ. Co., p.189
\reference{} Inagaki S., Saslaw W.C., 1985, ApJ, 292, 339
\reference{} Irwin M., Hatzidimitriou D., 1995, MNRAS, 277, 1354
\reference{} Klessen R.S., Kroupa P., 1998, ApJ, 498, 143
\reference{} Kroupa P., 1997, New Astronomy, 2, 139
\reference{} Kroupa P., 1998a, in Dynamics of Galaxies and Galactic Nuclei,
      Proc. Ser. I.T.A. no. 2, eds. W. Duschl \& C. Einsel, Heidelberg
      (astro-ph/9801047)
\reference{} Kroupa P., 1998b, in The Magellanic Clouds and Other
	Dwarf Galaxies, eds T. Richtler, J.M. Braun, Shaker Verlag, in
	press (astro-ph/9804255)
\reference{} Kroupa P., Tout C.A, Gilmore G., 1993, MNRAS, 262, 545
\reference{} Kuijken K., 1991, ApJ, 372, 125 
\reference{} Lynden-Bell D., 1976, MNRAS, 174, 695
\reference{} Madejski R., Bien R., 1993, A\&A, 280, 383
\reference{} Mihos J.C., Dubinski J., Hernquist L., 1998, ApJ, 494, 183
\reference{} Miller B.W., Whitmore B.C., Schweizer F., Fall S.M.,
	1997, AJ, 114, 2381
\reference{} Reid I.N., Hawley S.L., Gizis J.E., 1995, AJ, 110, 1838
\reference{} Scalo J., 1986, Fundam. Cosmic Phys., 11, 1
\reference{} Spitzer L., 1987, Dynamical Evolution of Globular
      Clusters, Princeton Univ. Press, Princeton 
\reference{} Syer D., Mao S., Mo H.J., 1998, MNRAS, preprint
       (astro-ph/9711160)
\reference{} Vallenari A., Bettoni D., Chiosi C., 1998, A\&A, 331, 506
\reference{} van den Bergh S., 1986, AJ, 91, 271
\reference{} Westerlund B.E., 1997, The Magellanic Clouds, Cambridge
	University Press, Cambridge
\reference{} Whitmore B.C., et al., 1998, in preparation
\reference{} Whitmore B.C., Schweizer F., 1995, AJ, 109, 960

\end{references}
\end{document}